\documentstyle[12pt]{article}

\begin{document}

\noindent

\def\med{{1\ov 2}}
\def\hepth#1{ {\tt hep-th/#1}}

\def\be{\begin{equation}}
\def\ee{\end{equation}}
\def\bes{\begin{equation*}}
\def\ees{\end{equation*}}

\def\beqa{\begin{eqnarray}}
\def\beqas{\begin{eqnarray*}}
\def\eeqa{\end{eqnarray}}
\def\eeqas{\end{eqnarray*}}
\def\bea{\begin{eqnarray}}
\def\eea{\end{eqnarray}}

\def\cl{\mbox{\tiny (class)}}

\def\etal{{\it et al.\/}}%--- et al.
\def\ie{{\it i.e.\/}}%--- i.e.
\def\eg{{\it e.g.\/}}%--- e.g.
\def\Cf{{\it Cf.\ }}%--- Cf.
\def\cf{{\it cf.\ }}%--- cf.

\def\al{\alpha}
\def\lam{\lambda}
\def\blam{\bar\lambda}
\def\th{\theta}
\def\bth{\bar\theta}
\def\bsigma{\bar\sigma}
\def\bpsi{\bar\psi}

\def\bu{{\bar 1}}
\def\bd{{\bar 2}}
\def\bt{{\bar 3}}
\def\bc{{\bar 4}}
\def\dgam{\dot\gamma}
\def\dal{\dot\alpha}
\def\dbet{{\dot\beta}}

\def\Re{{\rm Re}}
\def\Im{{\rm Im}}

\def\H{{\cal H}}
\def\tr{{\rm tr}}
\def\Tr{{\rm Tr}}
\def\F{{\cal F}}
\def\N{{\cal N}}

\def\d{\partial}
\def\ov{\over}
\def\bD{\bar D}
\def\pder#1#2{{{\partial #1}\over{\partial #2}}}%--- partial derivative
\def\der#1#2{{{d #1}\over {d #2}}}%--- full derivative
\def\ppder#1#2#3{{\partial^2 #1\ov\partial #2\partial #3}}
\def\dpder#1#2{{\partial^2 #1\ov\partial #2 ^2 }}
\def\bemat{\left(\begin{array}}
\def\enmat{\end{array}\right)}

\def\Fpk{\alpha\!\cdot\!\!\F'_k}
\def\Fpu{\alpha\!\cdot\!\!\F'_1}
\def\Fpd{\alpha\!\cdot\!\!\F'_2}
\def\Fppk{\alpha\!\cdot\!\!\F''_k\!\!\cdot\! \alpha}
\def\Fppu{\alpha\!\cdot\!\!\F''_1\!\!\cdot\! \alpha}
\def\Fppd{\alpha\!\cdot\!\!\F''_2\!\!\cdot\! \alpha}
\def\cdotsh{\!\cdot}

\begin{titlepage}
\begin{flushright}
{ ~}\vskip -1in
US-FT/1-99\\
\hepth{9901006}\\
Jan 1999\\
\end{flushright}

\vspace*{20pt}
\bigskip

\centerline{\Large STRONG COUPLING EXPANSION AND}
\bigskip
\centerline{\Large SEIBERG--WITTEN--WHITHAM EQUATIONS}
\vskip 0.9truecm
\centerline{\large\sc Jos\'e D. Edelstein\footnote{\tt
edels@fpaxp1.usc.es} and Javier Mas\footnote{\tt jamas@fpaxp1.usc.es}}

\vspace{1pc}

\begin{center}
{\em Departamento de F\'\i sica de Part\'\i culas, Universidade de Santiago 
de Compostela,\\
E-15706 Santiago de Compostela, Spain.} \\

\vspace{5pc}

{\large \bf Abstract}
\end{center}

We study the Seiberg-Witten-Whitham equations in the strong coupling regime of 
the ${\cal N}=2$ super Yang-Mills theory in the vicinity of the maximal
singularities.
In the case of $SU(2)$ the Seiberg-Witten-Whitham equations fix completely
the strong coupling expansion.
For higher rank $SU(N)$ they provide a set of non-trivial constraints on the 
form of this expansion.
As an example, we study the off-diagonal couplings at the maximal point for 
which we propose an ansatz that fulfills all the equations.

\end{titlepage}
\setcounter{footnote}{0}

The low energy dynamics of ${\cal N}=2$ supersymmetric gauge theories
is governed by a single holomorphic function ${\cal F}$ known as
the effective prepotential. 
A self-consistent proposal for this function has been done a few years 
ago by Seiberg and Witten \cite{SeiWitt} in the $SU(2)$ case and generalized 
to $SU(N)$ by many authors \cite{sun}.
The Seiberg-Witten solution for the effective theory of ${\cal N}=2$ super 
Yang-Mills can be embedded into the Whitham hierarchy associated to the 
periodic Toda lattice \cite{toda}. 
The link between both constructions is summarized in the statement that the 
prepotential of the ${\cal N}=2$ Yang-Mills theory corresponds to the 
logarithm of the Toda's quasiclassical tau function. 

On the Whitham hierarchy side, there is a family of new variables entering 
into the prepotential known as {\em slow times}.
They can be promoted to spurion superfields that softly break ${\cal N}=2$ 
supersymmetry down to ${\cal N}=0$ with higher Casimir perturbations 
\cite{emm}, something that might drive the theory to a vacuum placed in 
the neighborhood of the so-called Argyres-Douglas singularities \cite{ad}.
Aside from this physical motivation and its mathematical interest, the 
Seiberg-Witten-Whitham (SWW) correspondence also provides us with new 
techniques to 
explore the structure of the effective action of ${\cal N}=2$ theories. 
In Ref.\cite{ITEP}, for example, the second derivatives of the prepotential 
with
respect to the Whitham times were computed and shown to be given in terms
of Riemann Theta functions associated to the root lattice of the gauge group.
When evaluated in the Seiberg-Witten theory, these formulae were shown to 
provide a powerful tool to compute all instanton corrections to the effective 
prepotential of $SU(N)$ ${\cal N}=2$ super Yang--Mills theory from the 
one-loop contribution through a recursive relation \cite{emm}.

It is the aim of this letter to apply the Seiberg-Witten-Whitham equations to 
the strong coupling regime of the ${\cal N}=2$ super Yang-Mills theory with 
gauge group $SU(N)$.
We will start by considering ${\cal N}=2$ $SU(2)$ super Yang-Mills theory and 
show that, starting from the Whitham hierarchy side, the exact effective 
prepotential near the monopole singularity is obtained in a remarkably simple 
way up to arbitrary order in the dual variable.
We will then consider the generic $SU(N)$ case and show that the SWW equations 
do not give a closed recursive procedure to perform the same computation.
Nevertheless, as we will see, these equations can be used to study the 
couplings between different magnetic $U(1)$ factors at the maximal 
singularities of the moduli space.

~

Let us start with a brief description of our general framework.
The low-energy dynamics of $SU(N)$ ${\cal N}=2$ super Yang-Mills theory is 
described in terms of the hyperelliptic curve \cite{sun}
\be
y^2 = P^2(\lambda,u_k) - 4\Lambda^{2N},
\label{hyper}
\ee
where  $P(\lambda,u_k) = \lambda^{N} - \sum_{k=2}^N u_k \lambda^{N-k}$
is the characteristic polynomial of $SU(N)$ and $u_k,~ k=2,...,N$
are the Casimirs of the gauge group. 
This curve can be identified with the spectral curve of the $N$ site periodic 
Toda lattice and, moreover, the prepotential of the effective theory is 
essentially the logarithm of the corresponding quasiclassical tau function, 
hence depending 
on the slow times $T_n$, $1\leq n\leq N-1$, of the corresponding Whitham 
hierarchy\footnote{It was shown in Refs.\cite{emm,ITEP} that the relevant 
quantities are
not $T_n$ but the rescaled Whitham slow times $\hat T_n = T_n/T_1^n$ (in fact,
the SWW correspondence also requires a rescaling of the scalar variables $\hat 
a^i = T_1 a^i$ and, consequently, of the Casimirs $\hat u_k = T_1^k u_k$).
We will always work in what follows with the rescaled variables and, therefore,
hats will be omitted everywhere.} \cite{toda}. 
The derivatives of the prepotential with respect to the Whitham slow times 
have been obtained in Ref.\cite{ITEP}.
When restricted to the submanifold $T_{n>1}=0$, where the Seiberg-Witten 
solution lives (we may identify $T_1$ with $\Lambda$ in this submanifold 
\cite{emm}), the first derivatives are simply
\be
\pder{\F}{\log\Lambda\,} = {\beta\ov 2\pi i } \H_{2} ~~~~~~~~~~~~~ 
\pder{\F}{ T_n} {}~=~ {\beta\ov 2\pi i n}  \H_{n+1} ~, 
\label{firstder}
\ee
while the second order derivatives with respect to the Whitham slow times
result in \cite{emm}
\beqa
\dpder{\F}{(\log\Lambda)} & = & -{\beta^2\ov 2\pi i}
\pder{\H_{2}}{ a^i}\pder{\H_{2}}{ a^j}{1\ov i\pi}
\d_{\tau_{ij}}\log\Theta_E(0|\tau) ~, \nonumber \\
\ppder{\F}{\log\Lambda\,}{ T_n} &=& -{\beta^2\ov 2\pi i n}
\pder{\H_{2}}{ a^i}\pder{\H_{n+1}}{ a^j}{1\ov i\pi}
\d_{\tau_{ij}}\log\Theta_E(0|\tau) ~, \label{lasecudef} \\
\ppder{\F}{ T_m}{ T_n} &=& -{\beta\ov 2\pi i} \left(
\H_{m+1,n+1}+{\beta\ov mn}
\pder{\H_{m+1}}{ a^i}\pder{\H_{n+1}}{ a^j}{1\ov i\pi}
\d_{\tau_{ij}}\log\Theta_E(0|\tau) \right) ~, \nonumber
\eeqa
with $m,n\geq 2$ and $\beta= 2N$.
The functions $\H_{m,n}$ are certain homogeneous combinations of the
Casimirs $u_k$, given by
$$
\H_{m+1,n+1} = {N\ov mn} \hbox{res}_\infty \left(
P^{m/N}(\lambda)dP_+^{n/N}(\lambda) \right) ~~~ \mbox{and} ~~~
\H_{m+1} \equiv \H_{m+1,2} = u_{m+1} + {\cal O}(u_m) ~.
$$
Here $\left(\sum_{k=-\infty}^\infty c_k\lambda^k\right)_+ =
\sum_{k=0}^\infty c_k\lambda^k$ and $\hbox{res}_\infty$ stands for the 
usual Cauchy residue at infinity.
Notice that, for instance, $\H_{2,2} = \H_{2} = u_2,~\H_{3,2} =\H_{3}= u_3$ 
and $\H_{3,3} = u_4+ {N-2\over 2N}u_2^2$.

The characteristic $E$ of the Theta function entering the previous expressions
can be read, for example, from the blow-up 
formula derived in Ref.\cite{mm} for twisted ${\cal N}=2$ supersymmetric
gauge theory to be $\vec \alpha=(0, \dots, 0)$ and $\vec \beta = (1/2, 
\dots, 1/2)$.
This is the --even and half-integer-- characteristic of the Theta function
when we express it in terms of {\it electric} variables. 
If we are to consider the physics near the ${\cal N}=1$ (maximal) 
singularities, where generically $N-1$ magnetic monopoles become massless, 
we should use as local 
variables the dual $a_{D,k}$ reached from the former $a^k$ after an S duality 
transformation.
In Ref.\cite{emm}, a careful study of the $Sp(2r,{\bf Z})$ covariance of
Eqs.(\ref{lasecudef}) was performed.
In general, if we transform under an element $\Gamma$ of the duality group
${\rm Sp} (2r, {\bf Z})$, the variables $a^i$, $a_{D,i}$ transform as a vector 
$v^t =(a_{D,i},a^i)$, $v \rightarrow \Gamma v$, whereas the arguments $\xi$, 
$\tau$ of the Theta function change as follows:
\bea
\tau & \rightarrow & \tau^{\Gamma} = (A \tau + B) (C \tau + D)^{-1} ~, \\
\xi & \rightarrow & \xi^{\Gamma} = [ (C\tau + D)^{-1}]^t \xi ~,
\eea
with $A^t D-C^t B = 1_r$, $A^tC= C^t A$ and $B^tD= D^t B$. 
The characteristics (understood as row vectors) transform as
\bea
\alpha &\rightarrow & \alpha ^{\Gamma}= D\alpha-C\beta +{1 \over 2} {\rm
diag}(CD^t) ~, \\
\beta  &\rightarrow & \beta ^{\Gamma}= -B\alpha+A\beta +{1 \over 2} {\rm
diag}(AB^t) ~.
\label{chartrans}
\eea
From these equations, the characteristic of the Theta function near the 
${\cal N}=1$ points, which we will call D, can be computed to be 
$\vec\alpha=(1/2, \dots, 1/2)$ and $\vec\beta =(0, \dots, 0)$.
The second derivatives of the prepotential with respect to the Whitham
slow times can be written in this patch of the moduli space as in 
Eqs.(\ref{lasecudef}) by replacing $\tau\to\tau_D$ and using the dual
characteristic D in Riemann's Theta function.
These formulae should be suitable to study the physics of ${\cal N}=2$
super Yang-Mills theory near the ${\cal N}=1$ singularities. 
Let us first consider, for simplicity, the SU(2) case.

~

\noindent
\underline{\em SU(2)}
\vspace{1mm}

There are two ${\cal N}=1$ singularities related by the unbroken $Z_2$
subgroup of the gauge group, occuring at each of the two points of the 
vanishing locus ${\cal C}_\Lambda : \{u = \pm 2\Lambda^2\}$ (here
$u \equiv u_2 = \H_2$ is the quadratic Casimir of $SU(2)$).
They correspond to a massless monopole and a massless dyon, in a 
given symplectic basis of cycles on the torus.
We will study the point $u = 2\Lambda^2$: all quantities at the other
point can be obtained through a $Z_2$ transformation.
The generic form of the strong coupling expansion of the prepotential at
such singularity is given by:
\be
\F = {1\ov 4\pi i} a_{D}^2 \log \frac{a_{D}}{\Lambda} + {i \Lambda^2\ov 
2 \pi}\sum_{s=0}^\infty \F_s \left(\frac{ia_D}{\Lambda}\right)^s ~,
\label{prepsu2}
\ee
where the logarithmic term, coming from the one-loop diagram that involves
the light monopole, has the appropriate sign and factor making manifest 
that the theory is non-asymptotically free and that there is a monopole
hypermultiplet weakly coupled to the dual photon for $a_D\to 0$.
The remaining power series expansion comes from the integration of 
infinitely many massive BPS states.
We shall use in what follows the SWW correspondence to predict the 
strong coupling expansion of the prepotential (\ie, the values of the 
constants $\F_s$ in the expression above).
There is just one SWW equation in the $SU(2)$ case, namely the first one in 
({\ref{lasecudef}), and it reads
\be
\dpder{\F}{(\log\Lambda)} = {8 i\ov \pi}
\left(\pder{u}{ a_{D}}\right)^2 {1\ov i\pi}
\pder{}{\tau^D}\log\vartheta_2(0|\tau^D) ~,
\label{onlysu2}
\ee
where $\vartheta_2(0|\tau^D)$ is Jacobi's Theta function,
\be
\vartheta_2(0|\tau^D) = \sum_{n=-\infty}^{\infty}
e^{i\pi(n+1/2)^2\tau^D} ~.
\label{theta2}
\ee
Using the ansatz (\ref{prepsu2}), one can compute the derivative of the Theta 
function and accomodate the result as follows
\be
{1\ov i\pi} \pder{}{\tau^D}\log\vartheta_2(0|\tau^D) = 
\left[\sum_{n=0}^{\infty} \Xi(a_D)^{n(n+1)/2}\right]^{-1}
\left[\sum_{n=0}^{\infty} \left(n+\med\right)^2~\Xi(a_D)^{n(n+1)/2}\right] ~,
\label{step1}
\ee
where $\Xi(a_D)$ is given by
\be
\Xi(a_D) = e^{3/2} \left(\frac{a_D}{\Lambda}\right)
\prod_{s=2}^{\infty} 
\exp\left\{s(s-1)\F_s\left(\frac{ia_D}{\Lambda}\right)^{s-2}\right\} 
\approx e^{3/2} \left(\frac{a_D}{\Lambda}\right) +
{\cal O}\left(\frac{a^2_D}{\Lambda^2}\right) ~.
\label{aux1}
\ee
The expression above in terms of $\Xi(a_D)$ makes it simple to expand 
(\ref{step1}) up to arbitrary order in $a_D$.
It is interesting to mention that Eqs.(\ref{onlysu2})--(\ref{step1}) give
a non-trivial consistency check of the dual characteristic previously derived.
Indeed, being $a_D$ a local coordinate of the moduli space, its corresponding 
characteristic must be even and half--integer.
This fact, together with the extra $-1/2$ factor in the logarithmic term of the
prepotential (\ref{prepsu2}) --related to the fact that the massless magnetic 
monopole comes in an ${\cal N}=2$ hypermultiplet--, leads to a unique 
possibility 
that makes both members of Eq.(\ref{onlysu2}) to fit when an expansion in the 
dual variable is considered, this being $(\alpha = 1/2\,,\beta=0)$.
Still, in order to compute the RHS of (\ref{onlysu2}), we must use the first
equation in (\ref{firstder}) (which is nothing but the RG equation derived in 
Ref.\cite{rengr}) to obtain
\be 
\pder{u}{ a_{D}} = -\frac{1}{4}a_D + \frac{\Lambda^2}{4a_D}
\sum_{s=1}^\infty s(s-2)\F_s \left(\frac{ia_D}{\Lambda}\right)^s ~,
\label{ingr1}
\ee
while the LHS of the SWW equation is simply
\be
\dpder{\F}{(\log\Lambda)} = \frac{i\Lambda^2}{2\pi}
\sum_{s=0}^\infty (s-2)^2\F_s \left(\frac{ia_D}{\Lambda}\right)^s ~.
\label{ingr2}
\ee
These expressions can be inserted into Eq.({\ref{onlysu2}) and expanded
around the vanishing value of the dual variable. 
At the end, one can recursively compute all the $\F_s$ coefficients of the 
strong coupling expansion, the first few terms being
\beqa
\F & = & {1\ov 4\pi i} a_{D}^2 \log \frac{a_{D}}{\Lambda} + {i \Lambda^2\ov 
2 \pi} \left\{ -1 + 4\left(\frac{ia_D}{\Lambda}\right) -
\left(\frac{3}{4}+\med\log(16i)\right)\left(\frac{ia_D}{\Lambda}\right)^2
+ \frac{1}{16}\left(\frac{ia_D}{\Lambda}\right)^3 \right. \nonumber \\
& & \left. + \frac{5}{2^9}\left(\frac{ia_D}{\Lambda}\right)^4 + 
\frac{11}{2^{12}}\left(\frac{ia_D}{\Lambda}\right)^5 +
\frac{63}{2^{16}}\left(\frac{ia_D}{\Lambda}\right)^6 + 
\frac{527}{5\cdot 2^{18}}\left(\frac{ia_D}{\Lambda}\right)^7 +
{\cal O}(a_D^8) \right\} ~,
\label{resultsu2}
\eeqa
in precise agreement with the results of Ref.\cite{klemm}.
Notice that the solvability or completeness of the recursive relation that
results from this procedure is not granted.
It is also interesting to remark that the SWW correspondence allowed us to
obtain the exact prepotential near the strong coupling singularity {\em 
without} an explicit knowledge of the actual solution $(a(u),a_D(u))$.

~

\noindent
\underline{\em SU(N)}
\vspace{1mm}

The quantum moduli space of $SU(N)$ ${\cal N}=2$ super Yang--Mills theory has 
$N$ maximal singularities where $N-1$ monopoles become massless 
simultaneously. 
Physical quantities in the neighborhood of any of these singularities can be 
translated to a patch in the vicinity of any other by the action of the 
unbroken discrete subgroup ${\bf Z}_N$. 
We will consider in what follows the point where $\H_2$ is real and positive.
The strong coupling expansion of the prepotential at such singular point can 
be written in terms of appropriate $a_{D,i}$ variables as\footnote{We follow 
here the conventions of Ref.\cite{DS} to fix the first three terms of the 
expansion.}
\beqa
\F & = & \frac{N^2}{2\pi i}\Lambda^2 + 
\frac{2N\Lambda}{\pi}\sum_{k=1}^{N-1} a_{D,k}\,\sin(\pi{k}/N) + {1\ov 4\pi 
i}\sum_{k=1}^{N-1}a_{D,k}^2 \log \frac{a_{D,k}}{\tilde\Lambda_k} \nonumber \\
& & + \frac{1}{2} \sum_{k\neq l=1}^{N-1} \tau_{kl}^{\rm off}\,a_{D,k} a_{D,l}
+ {1\ov 2 \pi i}\sum_{s=1}^\infty \F_s(a_D)\Lambda^{-s} ~,
\label{elprep}
\eeqa
where $\F_s(a_D)$ are polynomials of degree $s+2$ in dual variables, 
$\tilde\Lambda_k = e^{3/2}\Lambda\sin\hat\theta_k$ (with $\hat\theta_k = \pi 
k/N$) and $\tau_{ij}^{\rm off}$, $i \not=j$ are the values 
of the off-diagonal entries of the coupling constant at the ${\cal N}=1$ point,
$a_{D,i}=0$. 

Again, we shall introduce this prepotential expansion into the SWW equations
(\ref{lasecudef}).
These equations allow us to relate the strong coupling expansion of homogeneous
combinations of higher Casimir operators $\H_m$ with that of the prepotential.
Indeed, inserting (\ref{firstder}) into (\ref{lasecudef}), we have
\be
\pder{\H_{m}}{\log\Lambda\,} = 
-\beta\pder{\H_2}{a_{D,i}}\pder{\H_{m}}{a_{D,j}} 
{1\ov i\pi}\d_{\tau_{ij}} \log \Theta_D(0|\tau) ~.
\label{qucas}
\ee
In fact, from Eq.(\ref{qucas}), both the full expansion of the prepotential 
around the semiclassical patch for $SU(N)$ ${\cal N}=2$ super Yang--Mills 
theory \cite{emm} as well as the strong coupling expansion near the maximal 
singularities in the $SU(2)$ case were obtained.
The generalization of this last result to $SU(N)$, however, faces some
difficulties.
First, the dual characteristic does not lead to the vanishing of some higher
contributions as it happens in the semiclassical expansion with the `electric'
characteristic E, a feature that allowed in this last case to explicitely show
recursivity \cite{emm}.
Second, the polynomials $\F_s(a_D)$ are not constrained by classical symmetries
--as the Weyl group in the semiclassical expansion-- being the amount of 
unknowns, then, considerably higher.
Third, the prepotential (\ref{elprep}) has powers of $\Lambda$ of both signs 
such that, in spite of the fact that a grading still exists, many higher terms 
of the expansion appear in the lowest equations spoiling 
recursivity\footnote{Notice that the same difficulties should also appear for 
$SU(2)$ though, as we have explicitely shown above, this is not the case. The
reason for this singular character of the gauge group $SU(2)$ is nothing but
the one-dimensional character of its Cartan subalgebra as a detailed analysis 
makes manifest.}.

The SWW equations do not seem to be instrumental to study the full strong 
coupling expansion of $SU(N)$ ${\cal N}=2$ super Yang--Mills theory.
Other methods have been derived in the literature to tackle the problem of
computing the higher threshold corrections $\F_s(a_D)$.
For example, in Ref.\cite{dHP} this has been accomplished by
parametrizing the neigborhood of the maximal singularities with a family
of deformations of the corresponding auxiliary singular Riemann manifold.
However, this formalism is not sensitive to quadratic terms in the 
prepotential. 
Thus, it does not give an answer for the couplings between different
magnetic $U(1)$ factors at the maximal singularities of the moduli space, 
encoded in the coefficients $\tau_{ij}^{\rm off}$.
The existence and importance of such terms has been first pointed out in
Ref.\cite{DS} by using a scaling trajectory that smoothly connects the
maximal singularities with the semiclassical region.
These terms also appear in the expression of the Donaldson--Witten
functional for gauge group $SU(N)$ \cite{mm}.
To our knowledge, a closed formula for these off-diagonal couplings has not
been obtained so far, except for the gauge group $SU(3)$ \cite{klemm}.
In the rest of this letter, we will apply the SWW equations to the
solution of this problem.

~

Let us first remark that Eq.(\ref{qucas}) is also valid for the higher
Casimirs $h_n = 1/n\,{\rm Tr}\phi^n$, $n=2,...,N$ \cite{ITEP}.
They, as well as their particular combinations encoded in $\H_n$, are 
homogeneous functions of $a_D$ and $\Lambda$ of degree $n$. 
Thus, at the ${\cal N}=1$ singularities, the LHS of eq.(\ref{qucas}) is 
simply \cite{DS}
\be
{\partial h_n  \over \partial\log\Lambda} = n h_n =
\sum_{k=1}^{N} (2\cos\theta_k)^n ~,
\label{LHS}
\ee
where we used the fact that the eigenvalues of $\phi$ are $\phi_k = 
2\cos\theta_k$ with $\theta_k = (k-1/2)\pi/N$.
The derivative of the Casimir operators with respect to the dual variables 
can be computed at the same point of the moduli space, by using the explicit 
representation of the curve in terms of the Chebyshev polynomials \cite{emm},
resulting in
\be
{\partial h_n  \over \partial a_{D,j}}= -2i \sum_{l=0}^{[n/2-1]}
{n-1 \choose l} \sin (n-2l-1)\hat\theta_j ~.
\label{derHsun}
\ee
Finally, the leading couplings at the maximal singularity can be easily 
obtained from the expansion (\ref{elprep}) to be
\be
\tau^D_{ij }= {1 \over 2\pi i } \log \left( {a_{D,i}\over \Lambda_i}  \right)
\delta_{ij} + \tau_{ij}^{\rm off} ~,
\label{latau}
\ee
where the logarithmic divergence is nothing but the expected running of the
coupling constant to the point where $N-1$ monopoles become massless.

The derivative of the Theta function $\Theta_D$ with respect to the period 
matrix has the following expression when evaluated at the ${\cal N}=1$ 
singularity
\be
{1\ov i\pi}\partial_{\tau^D_{ij}}\log \Theta_D(0,\tau_D) = \frac{1}{4} 
\left(\sum_{\xi^k=\pm 1}\exp{(i\frac{\pi}{4}\xi^l\tau_{lm}^{\rm 
off}\xi^m)}\right)^{-1} 
\sum_{\xi^k=\pm 1}\xi^i\xi^j\exp{(i\frac{\pi}{4}\xi^l\tau_{lm}^{\rm 
off}\xi^m)} ~.
\label{dlogsun}
\ee
Now, we can insert the results (\ref{LHS}), (\ref{derHsun}) and 
(\ref{dlogsun}) in equation (\ref{qucas}):
\beqa
\frac{1}{2N}\sum_{k=1}^{N} (2\cos\theta_k)^n & = & \sum_{l=0}^{[n/2-1]}
{n-1 \choose l} \sin\hat\theta_i \sin (n-2l-1)\hat\theta_j
\left(\sum_{\xi^k=\pm 1}\exp{(i\frac{\pi}{4}\xi^l\tau_{lm}^{\rm 
off}\xi^m)}\right)^{-1} \nonumber \\ & & ~~~~~~~~ \times 
\sum_{\xi^k=\pm 1}\xi^i\xi^j\exp{(i\frac{\pi}{4}\xi^l\tau_{lm}^{\rm 
off}\xi^m)} ~.
\label{final}
\eeqa
We have $N-1$ equations and $(N-1)(N-2)/2$ unknowns (the components of the 
symmetric matrix $\tau_{ij}^{\rm off}$). 
Thus, Eq.(\ref{final}) has predictive power in its own for $SU(3)$ and 
$SU(4)$. 
Indeed, we obtain for these two cases the following values:
\be
SU(3): ~~~~~ \tau_{12}^{\rm off} = i/\pi \log{2} ~, ~~~~~~~~~~~~~~~~~~~~~~~~~~~
\label{su3}
\ee
\be
SU(4): ~~~~~ \left\{ \begin{array}{l} \tau_{12}^{\rm off} = 
\tau_{23}^{\rm off} = - i/\pi \log(\sqrt{2}-1) \\
\tau_{13}^{\rm off} = i/\pi \log\sqrt{2} \end{array} \right. ~.
\label{su4}
\ee
Notice that our result for $SU(3)$ coincides with that of Ref.\cite{klemm} 
while the ones for $SU(4)$ have not been found previously.
For higher $SU(N)$, further ingredients would be necessary in order to obtain 
the off-diagonal couplings at the ${\cal N}=1$ singularity.
Instead, inspired by the findings in the last section of Ref.\cite{DS}, we 
propose the following ansatz for $\tau_{mn}^{\rm off}$:
\be
\tau_{mn}^{\rm off} = \frac{i}{\pi}~\frac{2}{N^2}\sum_{k=1}^{N-1} \sin k\hat\theta_m 
\sin k\hat\theta_n \sum_{i,j=1}^N \tau_{ij}^{(0)} \cos k\theta_i 
\cos k\theta_j ~,
\label{result}
\ee
with $\tau_{ij}^{(0)}$ being given by
\be
\tau_{ij}^{(0)} = \delta_{ij} \sum_{k\neq{i}}\log(2\cos\theta_i-2\cos\theta_k)^2
- (1-\delta_{ij}) \log(2\cos\theta_i-2\cos\theta_j)^2 ~.
\label{logs}
\ee
Unfortunately, we are not aware of the existence of an equivalent expression 
in the literature to compare with.
We have checked numerically that, with our ansatz (\ref{result})--(\ref{logs})
for the off-diagonal couplings, the SWW equations are satisfied up to SU(11).
There is a second check that we can do using results that do not rely on 
Whitham equations at all.
Douglas and Shenker showed that the matrix $\tau^D_{mn}$ at any point of the
scaling trajectory diagonalizes in the basis $\{\sin k\hat\theta_n\}$ with 
certain particular eigenvalues (see Eqs.(5.9)--(5.12) of Ref.\cite{DS}).
The couplings (\ref{result})--(\ref{logs}) satisfy this restrictive condition 
in the limit of the scaling trajectory ending at the maximal singularity.
As long as our solution (\ref{result})--(\ref{logs}) matches two very stringent
conditions coming from different places, we believe that it provides a faithful
answer for $\tau_{mn}^{\rm off}$ as well as a highly non-trivial test of the 
Seiberg-Witten-Whitham correspondence proposed in Refs.\cite{emm,ITEP}. 

Although we have focused on $SU(N)$, what we have presented should be 
generalizable to other cases. 
In this respect we point out that relations such as (\ref{qucas}) hold for all 
the simply laced algebras \cite{lostak}.
Many interesting problems are raised by our approach.
In particular, it would be of great interest to implement the SWW equations all
along the scaling trajectory introduced in Ref.\cite{DS}, to be able to compute
with them at intermediate patches between the maximal singularities and the
semiclassical region of the quantum moduli space.
We hope to report on this problem elsewhere.

~

We would like to thank Michael Douglas for correspondence and Marcos 
Mari\~no for useful comments and a critical reading of the manuscript.
The work of J.D.E. is supported by a fellowship of the Ministry of Education 
and Culture of Spain. 
The work of J.M. was partially supported by DGCIYT under contract PB96-0960.

%---------------- Bibliografia-------------------

\end{document}